\documentclass[%
 reprint,
 amsmath,amssymb,
 aps,
superscriptaddress]{revtex4-2}

\usepackage{graphicx}
\usepackage{dcolumn}
\usepackage{bm}

\usepackage{color}

\def\VDO{VO$_2$}
\def\IVDO{i-VO$_2$}
\def\MVDO{m-VO$_2$}

\def\degC{~^\circ{\rm C}}

\begin{document}

\preprint{APS/123-QED}



\title{Thermally switchable polarization manipulation and diodelike transmission in scalable, resonator-free, mid-infrared metasurfaces with vanadium-dioxide grids }

\author{Andriy E.~Serebryannikov} 
\affiliation{Division of Physics of Nanostructures, ISQI, Faculty of Physics, Adam Mickiewicz University,\\ 61-614 Poznan, Poland} 
\author{Akhlesh Lakhtakia} 
\affiliation{NanoMM–Nanoengineered Metamaterials Group, Department of Engineering Science and \\ Mechanics, 
Pennsylvania State University, University Park, Pennsylvania 16802, USA} 
\author{Ekmel Ozbay}
\affiliation{Nanotechnology Research Center (NANOTAM), Institute of Materials Sciences (UNAM), Department of Physics, and Department of Electrical Engineering, Bilkent University, 06800 Ankara, Turkey}

\date{\today}

\begin{abstract}
We conceptualized three-array scalable metasurfaces 
comprising only three thin strip grids and numerically demonstrated their characteristics in the mid-infrared spectral regime for switchable polarization manipulation and related asymmetric diodelike transmission (AT). A few or all of the grids were taken to be made of \VDO, a phase-change material. For each proposed metasurface,
two effective structures and, therefore, two different functionalities exist, corresponding to the metallic and insulating phases of \VDO. The achieved scenarios of functionality switching that depend on the \VDO~phase are shown to significantly depend in the way in which \VDO~is incorporated to the metasurface.  
Switchable bands of polarization manipulation are up to 40~THz wide. The AT band can be modulated when Fabry--Perot (anti-)resonances come into play. 
\end{abstract}

\maketitle
\section{Introduction}
%
%
Polarization manipulation and related asymmetric transmission (AT) achievable \textit{via} quasiplanar metamaterials constitute a topic of considerable attention  from the mid-2000's. Most researchers were initially focused on few-layer metasurfaces   
comprising coupled periodic arrays of (complementary) subwavelength resonators
\cite{AT1,fedotov2006asymmetric,mutlu2013one,huang2012asymmetric,yan2013compact,apl2015dielectric}.
Later, geometric-phase gradient metasurfaces were proposed
for polarization control \cite{mueller2017metasurface,luo2017transmissive}. 
Recently, metasurfaces and heterostructures enabling efficient manipulation by 
orbital angular momentum have been suggested \cite{zhang2021polarization}. 
Notably, a single-array structure can deliver polarization 
manipulation/conversion with AT for incident circularly polarized (CP) plane waves, but at least two coupled arrays are needed for AT when the incident plane wave is linearly polarized (LP). 
Mid-infrared (MIR) and near-infrared applications need rather simple design solutions to assure facile fabrication. In particular, polarization-converting, few-layer, resonator-free anisotropic structures comprising grids and related components have been suggested for different parts of electromagnetic spectrum \cite{pfeiffer2014high,MLAY02,breakingMalus}. Such structures are often considered to be dispersionless and, therefore, promising for applications. 

The research landscape for polarization control changed dramatically when tunable materials were introduced to metasurfaces and metamaterials, thereby 
enabling reconfigurable and multifunctional devices 
\cite{R1,maguid2017multifunctional,kumar2020tricontrolla}. 
Magnetically, electrically, thermally, and optically tunable materials such as graphene, InAs, InSb, GeTe, CdTe, \VDO, 
and ITO are often considered, in this context. In particular, 
\VDO~is a thermally tunable phase-change material useful from the terahertz 
(THz) to the visible spectral regimes \cite{cueff2020vo2,liu2018recent,lu2021metal,shi2019recent}. 
Transition from an insulator (I) phase to a metallic (M) phase, or from M to I phase, under the variation of a control parameter may occur for many natural materials~\cite{Imada}. 
One of them is vanadium dioxide (\VDO), an attractive phase-change material displaying a hysteretic insulator-metal-insulator transition~\cite{Adler}. 
It is underlaid by 
the monoclinic-to-tetragonal phase change that occurs as temperature, $T$ is raised from a value slightly lower than $58\degC$ to a value slightly above $72\degC$~\cite{Kakiuchida}. 
As a result, \VDO~behaves as a dissipative insulator (\IVDO) when monoclinic and metallic (\MVDO) when tetragonal, provided the free-space wavelength $\lambda \gtrsim 1300$~nm.

The simplest form in which \VDO~has been used in photonic heterostructures and metasurfaces is thin homogeneous layers \cite{s2,s3,s4}, 
but the variety of functionality-switching scenarios then is limited. 
\VDO~pads/inserts have been incorporated in metallic subwavelength resonators
for THz applications, especially for switchable polarization manipulation and AT \cite{i1,i5,liu2019thermally,pvo5}.   
However, the feasibility of such meta-atoms is questionable for MIR applications.
Meta-atoms comprising only \VDO~or 
another thermally tunable material have also been proposed, to exploit resonances in  \MVDO~ 
and/or \IVDO~
\cite{butakov2018switchable,pvoplus,serebryannikov2018thermally,VAC,kepic2021optically,YKVO22021tunable,AdP2018,Xing}. 
Patterned thin films of a phase-change material have been utilized either to directly enable tunability or indirectly tune a functionality enabled by other components~\cite{rev011,MetacanvasA38,hajian2019vo,A1,Andriy2022}.

In this paper, we conceptualize and numerically justify the potential of three-array metasurfaces comprising only \VDO~grids and Ag grids, 
which are capable of thermally switchable wideband 
polarization manipulation and diodelike AT for LP waves in the  MIR  regime.
The motivation for this study includes avoidance of subwavelength 
resonators (made of either a metal or a phase change material, or both) to prevent possible 
resonance enhancement of absorption.
The particular aims are to: 
(a) demonstrate the role of \VDO~phase change as enabler of the functionality-switching scenarios 
involving polarization conversion and AT;
(b) clarify the way of using how \VDO~
in the grids affects the particular switching scenarios; 
(c) demonstrate scalability of the proposed designs in  the MIR regime for wider applicability; 
(d) show that conduction and dissipation in
\MVDO~are suitable to replace a metal so that the capability for polarization manipulation remains high; and
(e) clarify the emergence and role of Fabry--Perot resonances in the resulting functionality and scalability. 
Geometrically, the studied structures remind ones proposed by others in \cite{pfeiffer2014high,breakingMalus}, in contrast to which a 
phase-change material is used by us both as a functionality enabling material and as a 
material for tunable components. 
Taking into account that the required thickness of \VDO~strips ranges only from 40 to 180~nm, 
and that \MVDO~is not such a good conductor as Cu or Ag, the capability of the structures 
comprising \VDO~grids for polarization manipulation is not evident and needs justification.
The studied metasurfaces are easy to fabricate by using the standard techniques. 
Maximization of AT contrast and on-off switching contrast are beyond the scope.


\section{Results and Discussion}
Metasurfaces of three types are considered, differing in the way in which \VDO~is used. 
Every metasurface was assumed to be illuminated by a normally incident LP plane wave.  
 CST Studio Suite \cite{CST2} was used to calculate the transmission
coefficients $\tau_{nm}^M$ and $\tau_{nm}^I$,   
where the subscripts $m\in\left\{x,y\right\}$ and $n\in\left\{x,y\right\}$, respectively, identify the direction of the 
electric field of the incident LP plane wave and the direction of the transmitted electric field in the far zone. Front-face 
  and back-face illuminations correspond to the  
propagation along the $-z$ and $+z$ directions, respectively.  
These cases are identified by the superscripts $\rightarrow$ and $\leftarrow$. 
The considered structure is Lorentz reciprocal \cite{EAB}, so 
$\vert\tau_{yy}^\rightarrow\vert=\vert\tau_{yy}^\leftarrow\vert$, $\vert\tau_{xx}^\rightarrow\vert=\vert\tau_{xx}^\leftarrow\vert$, 
$\vert\tau_{xy}^\rightarrow\vert=\vert\tau_{yx}^\leftarrow\vert$, and $\vert\tau_{yx}^\rightarrow\vert=\vert\tau_{xy}^\leftarrow\vert$. 
Asymmetry in transmission may be desirably strong when either (a)
$\vert\tau_{xy}^\rightarrow\vert\gg\vert\tau_{xy}^\leftarrow\vert$ and 
$\vert\tau_{xy}^\rightarrow\vert\gg\vert\tau_{xx}^\rightarrow\vert$, 
or (b) $\vert\tau_{xy}^\rightarrow\vert\ll\vert\tau_{xy}^\leftarrow\vert$ and 
$\vert\tau_{xx}^\leftarrow\vert\ll\vert\tau_{xy}^\leftarrow\vert$. 
The complex relative permittivity of \VDO~was 
calculated from available experimental data \cite{dicken2009frequency}. The Drude model of
the complex relative permittivity of Ag was used, i.e.,
$\varepsilon_r^{{Ag}}(\omega)=1-
\left(\omega_p^{Ag}\right)^2/[\omega(\omega+i\gamma)]$, where 
$\omega=2\pi{f}$ is the angular frequency and $f$ is the linear
frequency, the angular plasma frequency   $\omega_p^{Ag}=1.37\times10^{16}~\mbox{rad}~\mbox{s}^{-1}$, and the collision frequency is  $\gamma=2.73\times10^{13}~\mbox{rad}~\mbox{s}^{-1}$ \cite{ordal1985optical}.

\subsection{Case of the central grid made of \VDO }
\begin{figure}[ht!]
\centering
\includegraphics[width=8.4cm]{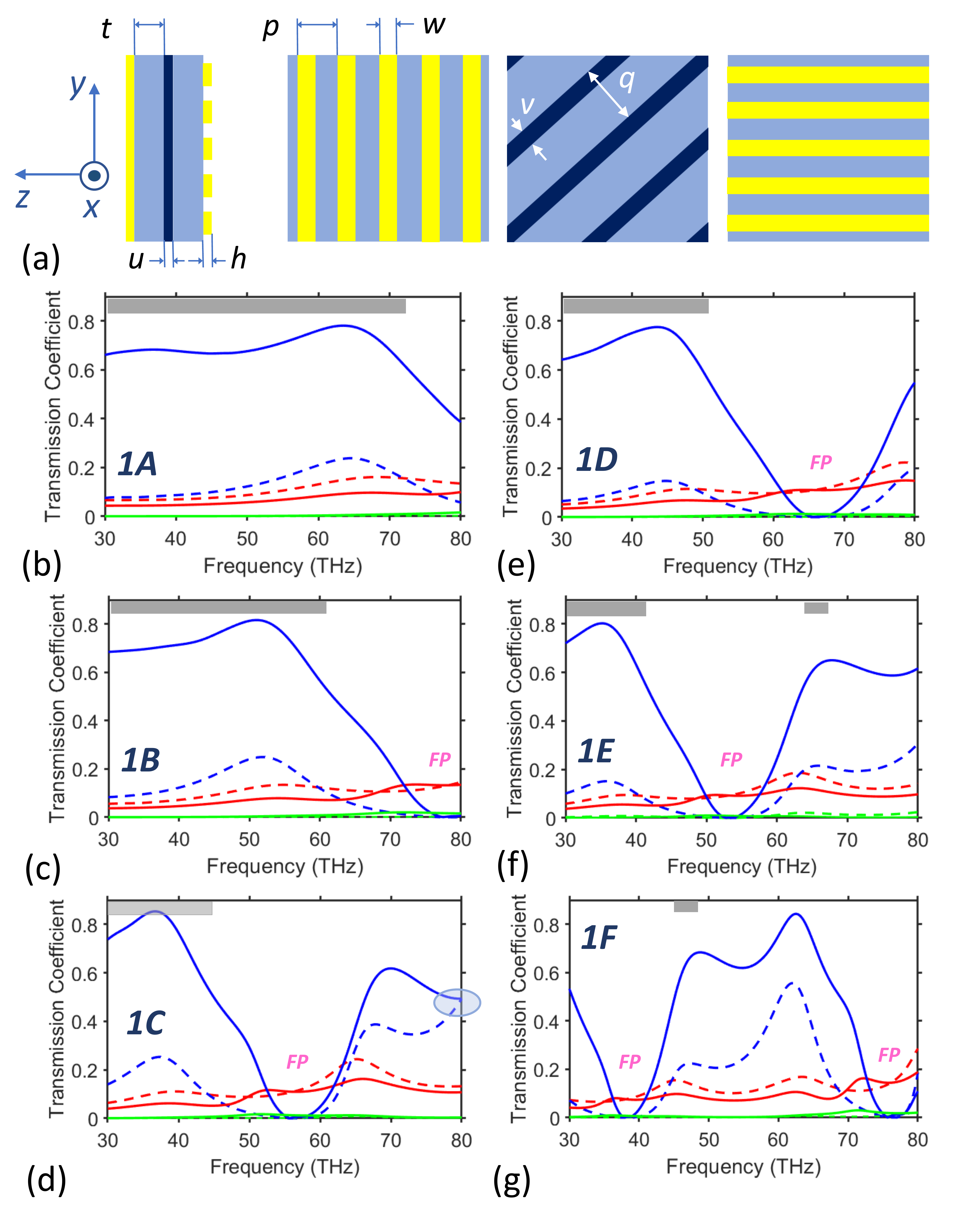}
\caption{Spectrums of transmission-coefficient magnitudes for six metasurfaces, 
1A to 1F, each comprising a \VDO~grid (shown in dark blue) separated from two Ag grids (yellow) by dielectric spacers (light blue).  
The central grid is oriented at $45^\circ$ with respect to the other two grids in the $xy$ plane. (a) From left to right: side cross-section view, front view, mid cross-section view, and back view.
(b)--(g) Transmission spectrums for spacer thickness  (b, c, d) $t=1000s$ nm and (e, f, g) $t=1500s$~nm, when relative permittivity of spacer $\varepsilon_{sp}=2.1$, $h=50s$~nm, $p=600s$~nm,  $w=p/2$, $q=\sqrt{2}p$, $u=2h$, and $v=w$; scaling coefficient (b,e) $s=1$, (c,f) 1.25, and (d,g) 1.75.  Solid lines: \MVDO, dashed lines: \IVDO; blue lines: $|\tau_{yx}^{\rightarrow}|$, 
green lines: $|\tau_{xy}^{\rightarrow}|$, and red lines: $|\tau_{xx}|=|\tau_{yy}|$. Grey rectangles 
indicate the $f$-ranges which may be suitable for switchable AT. }
\label{fig:FIG1}
\end{figure}

First, we consider the metasurfaces 1A--1F, each with a \VDO~grid separated from two Ag grids by dielectric spacers with the central grid oriented at $45^\circ$ in the transverse plane with respect to the other two grids, as shown in Fig.~\ref{fig:FIG1}(a). 
Thus, each metasurface has two conducting grids for \IVDO~and three conducting grids for \MVDO. 
The central grid works like a phase screen for \IVDO, as we have recently shown \cite{Andriy2022}. 
Accordingly, switching 
is possible between transmission with the dominant cross-polarized components 
for \MVDO~and reflection for \IVDO. 
Graphs of the  transmission-coefficient magnitudes vs.~$f$ are presented for all six
metasurfaces in Figs.~\ref{fig:FIG1}(b)--(g).
Let $\psi=|\tau_{yx}^{M\rightarrow}|^2/|\tau_{xx}^{M\rightarrow}|^2$ and $\zeta=(|\tau_{yx}^{M\rightarrow}|^2+|\tau_{xx}^{M\rightarrow}|^2)/(|\tau_{yx}^{I\rightarrow}|^2+|\tau_{xx}^{I\rightarrow}|^2)$. 
Switchable AT appears, for instance, at $f=64$ THz in Fig.~\ref{fig:FIG1}(b) for metasurface 1A, where $\psi=87$ and $\zeta=8.2$, and at $f=37$ THz in Fig.~\ref{fig:FIG1}(e) for metasurface 1D, where $\psi=147$ and $\zeta = 10$.
Similar transmission features are observed in Figs.~\ref{fig:FIG1}(b)--(d) for metasurfaces
1A--1C  as in  Figs.~\ref{fig:FIG1}(e)--(g) for metasurfaces
1D--1F, the spacers in the first three metasurfaces being two-thirds in thickness compared to 
those in the last three metasurfaces.

As metasurfaces 1A and 1D differ only in spacer thickness $t$, their transmission features are \textit{mutually scalable},
albeit approximately; the same statement can be made about metasurfaces 1B and 1E, as well as about metasurfaces
1C and 1F. This scaling can be approximately quantified by comparing the spectral locations of the 
typical minimums and maximums. 
Spectral features are downshifted and compressed in the $f$-domain when the scaling coefficient $s$ is increased. For instance, the minimum of $|\tau_{yx}^{M\rightarrow}|$ occurs at 65 THz for 1D [Fig.~\ref{fig:FIG1}(e)], 53 THz for 1E [Fig.~\ref{fig:FIG1}(f)], and 38 THz for
1F [Fig.~\ref{fig:FIG1}(g)], in inverse proportion to the spacer thickness.
 In contrast with conventional Fabry--Perot anti-resonances which appear for a homogeneous dielectric slab, all minimums of $\vert\tau_{yx}^{M\rightarrow}\vert$ in Figs.~\ref{fig:FIG1}(b)--(g) correspond to the condition $2\varepsilon_{sp}^{1/2}kt\approx{\ell}\pi$, where $\varepsilon_{sp}$ is the relative permittivity of the 
spacer, $k=2\pi/\lambda$ is the free-space wavenumber, and $\ell\in\left\{1,2\right\}$. These minimums, identified by \textit{FP} in the graphs,
 have the same spectral locations as the maximums in the case of a homogeneous slab dielectric slab sandwiched  by the two parallel Ag grids, i.e., when the back-face grid is not 
 rotated with respect to front-face grid. 
 Notably, most of the energy of the  incident plane wave is reflected at the minimums of $\vert\tau_{yx}^{M\rightarrow}\vert$, while reflections are weak at the maximums of $\vert\tau_{yx}^{M\rightarrow}\vert$.
%
%
%
By adjusting the spacer thickness, we can obtain on-off switchable, diodelike AT with a desired bandwidth. 

Incidentally, $|\tau_{yx}^{\rightarrow}|$ can be insensitive to the 
phase of \VDO~at a particular frequency,  e.g., $|\tau_{yx}^{I\rightarrow}| = |\tau_{yx}^{M\rightarrow}|$~at 80 THz in Fig.~\ref{fig:FIG1}(d) for 1C (shown by a bluish ellipse). 
On the other hand, different maximums of $|\tau_{yx}^{M\rightarrow}|$ show different sensitivity to the crystallographic phase of \VDO. 
For instance, in Fig.~\ref{fig:FIG1}(g) for 1F, $\zeta\approx10$ at 47 THz but $\zeta\approx2$ at 62 THz.

\subsection{Case of the outer grids made of \VDO }
\begin{figure}[ht!]
\centering
\includegraphics[width=8.4cm]{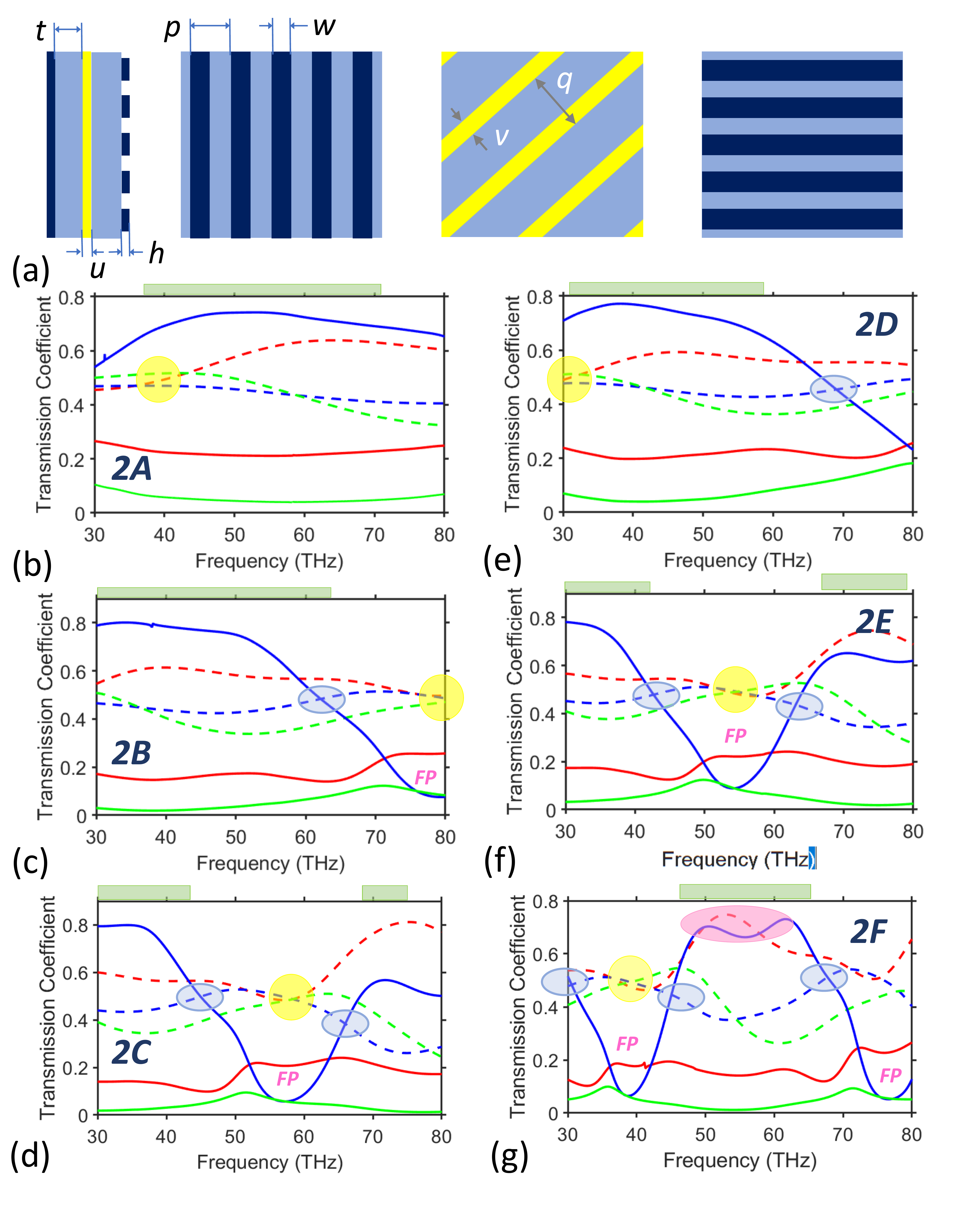}
\caption{Same as Fig.~\ref{fig:FIG1} but for metasurfaces 2A--2F, each comprising an Ag~grid (shown in yellow) separated from \VDO~grids (dark blue) by dielectric spacers (light blue). Scaling coefficient  (b,e) $s=0.75$, (c,f) 1.25, and (d,g) 1.75.  Light green rectangles in each plot  indicate the $f$-ranges which may be suitable for
 AT when \MVDO~is used.  
 }
\label{fig:FIG2}
\end{figure}

Next, let us swap Ag and \VDO~in metasurfaces 1A--1F to form
metasurfaces 2A--2F, each with an Ag~grid separated from two \VDO~grids by dielectric spacers, as shown in Fig.~\ref{fig:FIG2}(a). In this case, each metasurface has one conducting grid for \IVDO~and three conducting grids for \MVDO.
The front-face and the back-face grids work like a phase screen for \IVDO ~\cite{Andriy2022}. 
For \MVDO, the cross-polarized component of the transmitted plane wave is dominant compared to the co-polarized one, i.e.,~$\psi>10$. 
In particular, $\psi=11.7$ at 50 THz in Fig.~\ref{fig:FIG2}(b) for 2A, 19.7 at 30 THz in Fig.~\ref{fig:FIG2}(c) for 2B, and 28.4 at 62 THz in Fig.~\ref{fig:FIG2}(g) for 2F. 
For \IVDO, none of the transmission components is suppressed, i.e., they all are of the same order of magnitude. Thus, transmission mode is conserved when the crystallographic phase of \VDO~is thermally altered in 2A--2F, this  characteristic  being absent in Fig.~\ref{fig:FIG1} for 1A--1F. 
Moreover, spectral regimes with 
$|\tau_{yx}^{I\rightarrow}|\approx|\tau_{xx}^{I\rightleftharpoons}|\approx|\tau_{xy}^{I\rightarrow}|\approx|\tau_{yy}^{I\rightleftharpoons}|$ are observed (shown by  yellowish  circles). 
Then, both co- and cross-polarized transmission channels contain \textit{equal energy}, regardless of whether the front face or the back face is illuminated. It is evident at $f=80$~THz in Fig.~\ref{fig:FIG2}(c) for 2B and at $f=55$~THz in Fig.~\ref{fig:FIG2}(f) for 2E. 

The features observed in Figs.~\ref{fig:FIG2}(e)--(g) are similar to the ones in 
Figs.~\ref{fig:FIG2}(b)--(d), so that 
a desired modulation of the transmission spectrum can be adjusted via 
variations in $t$ by means of Fabry--Perot resonances and anti-resonances, as well as a desired location of the AT band for \MVDO~and the equal-energy regime for \IVDO. These bands may coincide or not coincide, depending on which functionality-switching scenario is needed. 
Scaling manifests itself in Fig.~\ref{fig:FIG2}, similarly to Fig.~\ref{fig:FIG1}, in the redshift of transmission features with increase of the scaling coefficient $s$.
In addition, transmission spectrum becomes more compressed when a thicker spacer is used.
Despite the  structures and functional capabilities of the metasurfaces 2A--2F  being different from those of the metasurfaces 1A--1F,
the insensitivity of the latter to the crystallographic phase of \VDO~is also exhibited by the former for specific spectral regimes and field components (bluish ellipses). This is exemplified by $|\tau_{yx}^{\rightarrow}|$, e.g., at $f=62$ THz in Fig.~\ref{fig:FIG2}(c) for 2B and 43 THz as well as 63 THz in Fig.~\ref{fig:FIG2}(f) for 2E. The number of the related spectral regimes, 
in which the insensitivity is observed within the considered $f$-range, increases with $t$, since the transmission spectrum thereby becomes more compressed as the spacer thickens.

Among the switching regimes achievable with 2A--2F, the one observed in Fig.~\ref{fig:FIG2}(g) for $f\in[45,60]$~THz
should be mentioned. Here, the strongest transmission component has nearly the same magnitude, but represents either the co- or the cross-polarized component for the same incident plane wave, depending on which crystallographic phase of \VDO~is used (rosy  ellipse).
%
\begin{figure}[ht!]
\centering
\includegraphics[width=8.0cm]{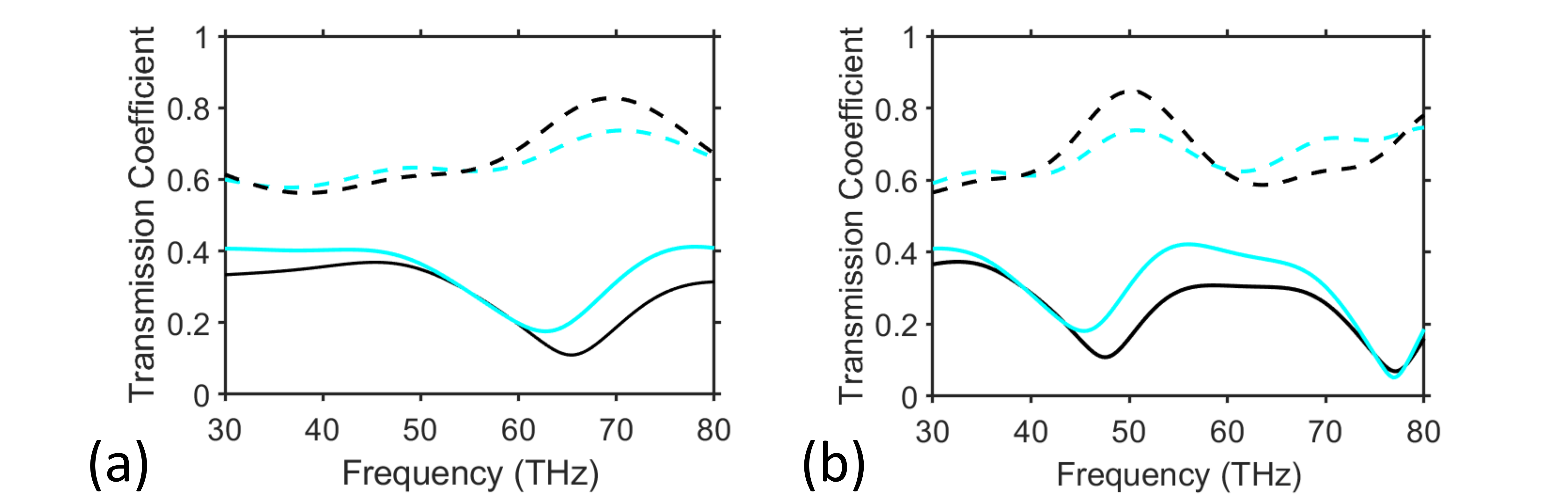}
\caption{Linear-to-circular polarization conversion for 
metasurfaces  
(a) 2C and (b) 2F.
Dashed cyan lines:
$|\tau_{Rx}^{\rightarrow}|$, 
solid cyan lines:
$|\tau_{Lx}^{\rightarrow}|$,  
dashed black lines:
$|\tau_{Ry}^{\rightarrow}|$, and
solid black lines:   
$|\tau_{Ly}^{\rightarrow}|$. 
 }
\label{fig:FIG3}
\end{figure}

For \IVDO, the metasurfaces 2A--2F can efficiently transmit an incident LP wave as a plane wave whose polarization state is close to being circular. We have adopted the usual optics convention: an LCP (left-handed CP) wave satisfies the condition 
$\nabla\times\vec{E}= k\vec{E}$, whereas an RCP (right-handed CP) wave satisfies the condition
$\nabla\times\vec{E}= -k\vec{E}$. Figure~\ref{fig:FIG3} presents transmission results for metasurfaces 2C and 2F, but the transmission 
is quantified now for CP outgoing waves. As observed in Fig.~\ref{fig:FIG3},  
both
$|\tau_{Rx}^{\rightarrow}|/|\tau_{Lx}^{\rightarrow}|>1$ and $|\tau_{Ry}^{\rightarrow}|/|\tau_{Ly}^{\rightarrow}|>1$ 
for both selected metasurfaces. Moreover, spectral regimes exist in which $|\tau_{Rx}^{\rightarrow}|/|\tau_{Lx}^{\rightarrow}|>5$ and/or $|\tau_{Ry}^{\rightarrow}|/|\tau_{Ly}^{\rightarrow}|>5$, so that linear-to-(nearly-)circular polarization conversion (LCPC) takes place.
%
The selective coupling of the incident LP waves to the outgoing CP waves allows the AT ranges 
for \MVDO~and LCPC ranges 
for \MVDO~to overlap when $s$ and $t$ are properly adjusted. 
In this case, change of phase of \VDO~results in bifunctional operation at fixed $f$. Otherwise, bifunctionality is obtained while using different $f$-ranges for 
the two phases of \VDO.

\subsection{Case of a metal-free structure}
\begin{figure}[ht!]
\centering
\includegraphics[width=8.0cm]{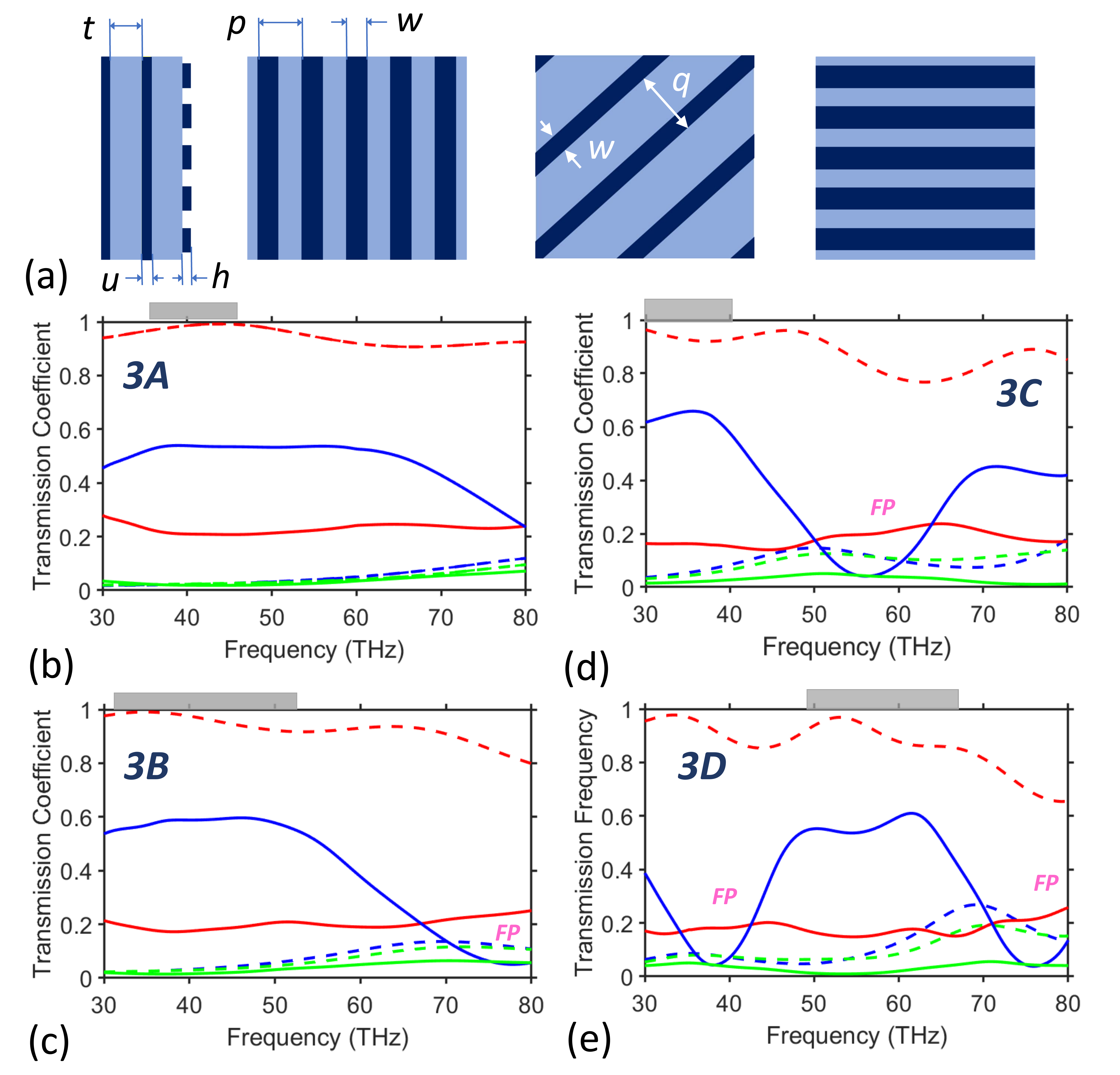}
\caption{Spectrums of transmission-coefficient magnitudes calculated for four metasurfaces, labeled 3A to 3D, each comprising three \VDO~grids (shown in dark blue) separated by dielectric spacers (light blue). 
(a) From left to right: side cross-section view, front view, mid cross-section view, and back view, respectively. (b)--(e) Transmission spectrums for spacer thickness  (b, c, d) $t=1000s$ nm and (e) $t=1500s$~nm. Scaling coefficient (b) $s=0.75$, (c) 1.25 and (d,e) 1.75. See Fig.~\ref{fig:FIG1} for other details. 
 }
\label{fig:FIG4}
\end{figure}

Finally, we consider the case of metal-free metasurfaces, i.e., when all three grids are made of \VDO. 
As a result, every metasurface in Fig.~\ref{fig:FIG4} has three conducting grids for \MVDO~but comprises only
  insulators for \IVDO. Accordingly, we obtain AT for \MVDO~and pre-dominantly co-polarized, weakly attenuated transmission for \IVDO. This scenario 
  of functionality switching is totally new as compared to 
  the ones in Figs.~\ref{fig:FIG1} and \ref{fig:FIG2}, and this is the main reason why we want to abandon metals. The results are presented in Figs.~\ref{fig:FIG4}(b)--(e) for the same geometric parameters as for Figs.~\ref{fig:FIG1}(b), (c), (d), and (g). 
%

The value of $\psi$ in Fig.~\ref{fig:FIG4} typically does not exceed 10. For instance, $\psi=6.5$ at 38 THz in Fig.~\ref{fig:FIG4}(b) for 3A, $\psi=7$ at 38 THz in Fig.~\ref{fig:FIG4}(c) for 3B,
and $\psi=12$ at 62 THz in Fig.~\ref{fig:FIG4}(e) for 3D. 
Along with the achievable magnitude of $|\tau_{yx}^{M\rightarrow}|$, this result 
is quite good since 
the metasurfaces do not contain any metal or cylindrical or other volume dielectric resonators  
(in contrast, say, with Ref.~\citenum{kepic2021optically}). 
Also for these metasurfaces 
we obtain scaling as the redshift of the basic transmission features occurring with increase of $s$, as observed from a comparison of Figs.~\ref{fig:FIG4}(b), \ref{fig:FIG4}(c) and \ref{fig:FIG4}(d), 
in spite of \VDO~being dispersive. 
Notably, the shifts of the minimums of $|\tau_{yx}^{M\rightarrow}|$ and the edges of the strong transmission ranges occur due to change 
of either $s$ or $t$, 
so these parameters provide two degrees of freedom for designing the transmission spectrum.\\

\section{Conclusion}
To summarize, we studied transmission through few-array metasurfaces comprising only three strip grids and exhibiting switchable polarization manipulation and related AT for LP 
plane waves in the 30--80-THz frequency range. The proposed structures are approximately scalable 
within this spectral regime. 
Subwavelength resonators, inserts, and pads  
are not needed that is important for ease of fabrication at MIR frequencies.
Diverse scenarios of functionality switching can be achieved, depending on whether 
\VDO~is used as the material of only the two outer, only the central, or all three grids. In all scenarios, 
well-pronounced AT connected with polarization manipulation occurs for \MVDO. 
It is switched for another functionality when \MVDO~is changed to \IVDO. 
This is true also when all three grids are made of \VDO, i.e., the metasurface is metal-free. 
Fabry--Perot (anti-)resonances manifest themselves in a way different from the case of a homogeneous dielectric slab of the same of thickness as two spacers in the studied structures and enable efficient modulation of the polarization effects and transmission. 
The observed capability for polarization manipulation and AT remains while adding 
a semi-infinite dielectric substrate at the back side (results are not shown). 
Besides the studied metasurfaces, those with only the front-face grid 
made of \VDO~were examined (results are not shown). 
The transmission efficiency can further be increased by using wider strips of the central 
grid than in the presented examples. 
For instance, $|\tau_{yx}^{M\rightarrow}|\approx0.8$ and 0.7 were obtained in the AT bands for the metasurfaces similar to 1F and 3D, respectively.



\begin{acknowledgments}
This work was supported by Narodowe Centum Nauki, Project UMO-2020/39/I/ST3/02413. 
A.~L.~thanks the Charles Godfrey Binder Endowment at the Pennsylvania State University for ongoing support. E.~O.~thanks the Turkish Academy of Sciences for partial support.
\end{acknowledgments}

%
%

\nocite{*}

\bibliography{apssamp}

\end{document}